\documentclass[11pt,epsf]{article}

\long\def\comment#1{\ifdim\overfullrule>0pt{\sf[{#1}]}\fi}

\usepackage{latexsym}
\usepackage{amsmath}
\usepackage{epsfig}
\usepackage{bbm}
\usepackage[usenames]{color}

\newtheorem{theorem}{Theorem}

 \newtheorem{lemma}{Lemma}

\newtheorem{corollary}[theorem]{Corollary}

\newcommand{\qed}{\hspace*{\fill}\mbox{$\Box$}}

\begin{document} \bibliographystyle{alpha}
\def\proofend{\hfill$\Box$\medskip}
\def\Proof{\noindent{\bf Proof:\ \ }}

\def\Sketch{\noindent{\bf Sketch:\ \ }}
\def\eps{\epsilon}

\title{A Counterexample to Beck's Conjecture on the Discrepancy of
  Three Permutations}

\author{Alantha Newman\thanks{\tt{alantha@dimacs.rutgers.edu}} \and
  Aleksandar Nikolov\thanks{\tt{anikolov@cs.rutgers.edu}}}

\maketitle

\begin{abstract}
Given three permutations on the integers 1 through $n$, consider the
set system consisting of each interval in each of the three
permutations.  J\'ozsef Beck conjectured (c.~1987) that the
discrepancy of this set system is $O(1)$.  We give a counterexample to
this conjecture: for any positive integer $n = 3^k$, we exhibit three
permutations whose corresponding set system has discrepancy
$\Omega(\log{n})$.  Our counterexample is based on a simple recursive
construction, and our proof of the discrepancy lower bound is by
induction.  This example also disproves a generalization of Beck's
conjecture due to Spencer, Srinivasan and Tetali, who conjectured that
a set system corresponding to $\ell$ permutations has discrepancy
$O(\sqrt{\ell})$~\cite{SST}.
\end{abstract}

\section{Introduction}

Given three permutations on the integers 1 through $n$, consider the
set system consisting of each interval in each of the permutations.
J\'ozsef Beck conjectured that there is always a coloring $\chi:[n]
\rightarrow \{-1, +1\}$ such that, after fixing this coloring, the
absolute value of any set in this set system is $O(1)$.  In other
words, he conjectured that the discrepancy of this set system is
$O(1)$.

We give a counterexample to this conjecture.  In particular, for each
integer $k > 0$, we give an instance of three permutations on the
ground set 1 through $3^k$ such that the discrepancy is at least
$\lceil k/3 + 1 \rceil$.  Setting $n= 3^k$, this yields a set of three
permutations with discrepancy at least $\lceil (\log_3{n})/3 + 1
\rceil$.

\subsection{Background}

The earliest reference to this conjecture that we have found is on
page 42 of the 1987 edition of Spencer's ``Ten Lectures on the
Probabilistic Method''~\cite{spencer-lectures}.  He presents a clever
proof that the discrepancy of two permutations is at most two, states
the conjecture for three permutations, and offers \$100 for its
resolution.  In the 1994 edition, Spencer attributes this conjecture
to Beck.  In a more recent book, Matou{\v{s}}ek says (on page 126)
that resolving Beck's conjecture ``remains one of the most tantalizing
questions in combinatorial discrepancy''~\cite{matousek2010geometric}.

Citing Beck's conjecture as motivation, Bohus showed that a
set system based on $\ell$ permutations always has
discrepancy $O(\ell \log{n})$~\cite{DBLP:journals/rsa/Bohus90}.  This
was later improved by Spencer, Srinivasan and Tetali who show that such a
set system actually has a coloring with discrepancy $O(\sqrt{\ell}
\log{n})$~\cite{SST}.  While Bohus gives an efficient algorithm to
find a coloring matching his upper bound, Spencer et al. leave open
the question of whether their bound can be achieved by an efficient
algorithm.  Since these latter results are via the entropy method, it
is possible that a constructive algorithm can be obtained via the
recent methods of Bansal, who gives constructive algorithms for
finding low discrepancy colorings for general set
systems~\cite{DBLP:conf/focs/Bansal10}.  Our results show that the
bounds of Bohus and of Spencer et al. are tight up to the factor
containing the number of permutations, $\ell$, i.e. these upper bounds
are tight for set systems based on a fixed number of permutations.
Spencer et al. also generalize Beck's conjecture positing that any set
system based on $\ell$ permutations has discrepancy
$O(\sqrt{\ell})$~\cite{SST}.

Recently, Eisenbrand, P{\'a}lv{\"o}lgyi and Rothvo{\ss} made a
surprising connection between Beck's conjecture and a well known open
question involving the additive integrality gap of the Gilmore-Gomory
LP relaxation for Bin Packing.  Specifically, they show that if Beck's
conjecture holds, then an optimal integral solution for the Bin
Packing problem is at most the optimal value of the LP relaxation plus
$O(1)$~\cite{DBLP:conf/soda/EDR11}.  They leave open the question of
whether a reduction in the other direction can be established: Does an
upper bound of $OPT_{LP} + O(1)$ on the size of an optimal integral
solution for the Bin Packing imply an $O(1)$ upper bound on the
discrepancy of three permutations?  In light of our results, such a
reduction would disprove the long-standing conjecture that the value
of an integral solution is upper bounded by $OPT_{LP} + O(1)$.  The
best known upper bound for the Bin Packing problem is $OPT_{LP} +
O(\log^2{n})$, which follows from a rounding procedure due to
Karmarker and Karp for the aforementioned LP
relaxation~\cite{DBLP:conf/focs/KarmarkarK82}.

\subsection{Basic Definitions and Notation}

Recall that for a set system ${\cal{S}} = \{S_1, S_2, S_3, \dots
S_m\}$ the discrepancy of the set system is:
\begin{eqnarray}
\text{disc}({\cal{S}}) & = & \min_{\chi} \max_{j \in [m]} |\sum_{i
  \in S_j} \chi(x_i)|.
\end{eqnarray}

Let $[n]$ denote the set of integers from 1 through $n$, and let
$[x,y]$ (where $x < y$) denote all integers from $x$ through $y$.  For
a coloring $\chi:[n] \rightarrow \{-1,+1\}$, if $S \subseteq [n]$, let
$\chi(S) = \sum_{j \in S} \chi(j)$.  We will usually use $n$ to denote
the length of the permutations, i.e. $n = 3^k$ for some specified
integer $k > 0$.

For some fixed $k$, the corresponding three permutations described in
Section \ref{sec:construction} will be denoted by $\pi_1^k, \pi_2^k$
and $\pi_3^k$.  Let $\alpha_i^k(x)$ denote the elements in positions 1
through $x$ in the permutation $\pi_i^k$, where $x \in [0,n]$.  In
other words, $\alpha_i^k(x)$ is a prefix of $\pi_i^k$ of length $x$.
Note that $\alpha_i^k(0)$ represents the empty set.  Given the three
permutations $\pi_1^k, \pi_2^k$ and $\pi_3^k$, the set system
${\cal{S}}_k$ consists of all sets $\alpha_i^k(x)$ for $x \in
[3^k]$.

We will also use the notion of sets corresponding to suffixes of the
permutations, even though these sets do not appear in our set systems.
Let $\omega_i^k(x)$ denote the elements in positions $x$ through $3^k$
in the permutation $\pi_i^k$, where $x \in [3^k+1]$.  In other
words, $\omega_i^k(x)$ is a suffix of $\pi_i^k$ of length $3^k-x+1$.
We define $\omega_i^k(3^k+1)$ to be the empty suffix.

\section{Recursive Construction}\label{sec:construction}

We give a construction for three permutations on the integers 1
through $n$, where $n=3^k$ for some integer $k > 0$.  Consider the
following recursive construction of three lists:
\begin{center}
\begin{tabular}{ccc}
$A$ & $B$ & $C$\\
$C$ & $A$ & $B$\\
$B$ & $C$ & $A$,
\end{tabular}
\end{center}
where $A$ represents the interval $[1,n/3]$, $B$ the interval $[n/3+1,
  2n/3]$, and $C$ the interval$[2n/3 + 1, n]$.  Each of the three
copies of $A$ (and $B$ and $C$, respectively) is divided further into
three equal sized blocks of consecutive elements, and these three
blocks are permuted as in the above construction.  This process of
dividing the blocks into three equal sized blocks and permuting them
according to the above construction is iterated $k$ times.  To
illustrate these actions, when $n=9$, this construction results in the
following three permutations:
\begin{center}
\begin{tabular}{ccccccccc}
1 & 2 & 3 & 4 & 5 & 6 & 7 & 8 & 9\\
9 & 7 & 8 & 3 & 1 & 2 & 6 & 4 & 5\\
5 & 6 & 4 & 8 & 9 & 7 & 2 & 3 & 1.
\end{tabular}
\end{center}
When $n=27$, the three permutations are:
\begin{tiny}
\begin{center}
\begin{tabular}{ccccccccccccccccccccccccccc}
1 & 2 & 3 & 4 & 5 & 6 & 7 & 8 & 9 &
10 & 11 & 12 & 13 & 14 & 15 & 16 & 17 & 18 &
19 & 20 & 21 & 22 & 23 & 24 & 25 & 26 & 27 \\
27 & 25 & 26 & 21 & 19 & 20 & 24 & 22 & 23 & 9 & 7 & 8 & 3 & 1 & 2 & 6
& 4 & 5 & 18 & 16 & 17 & 12 & 10 & 11 & 15 & 13 & 14 \\
14 & 15 & 13 & 17 & 18 & 16 & 11 & 12 & 10 & 23 & 24 & 22 & 26 & 27 &
25 & 20 & 21 & 19 & 5 & 6 & 4 & 8 & 9 & 7 & 2 & 3 & 1.
\end{tabular}
\end{center}
\end{tiny}

\subsection{Formal Description Based on Tensor Products}

We define the following three $3 \times 3$ matrices:
\begin{eqnarray*}
M_1 ~ = ~
\begin{pmatrix}
1 & 0 & 0 \\
0 & 1 & 0 \\
0 & 0 & 1
\end{pmatrix}, \quad
M_2 ~ = ~
\begin{pmatrix}
0 & 0 & 1 \\
1 & 0 & 0 \\
0 & 1 & 0 
\end{pmatrix}, \quad
M_3 ~ = ~
\begin{pmatrix}
0 & 1 & 0 \\
0 & 0 & 1 \\
1 & 0 & 0 
\end{pmatrix}.
\end{eqnarray*}
To construct an instance in which each permutation has size $3^k$,
define the three permutation matrices $M_i^{\otimes k}$ for $i \in
\{1,2,3\}$.  Each permutation matrix can be used to permute the
identity vector resulting in the three permutations in our
construction.  In other words, let {${\bf{v}}^k$} denote the column
vector of length $3^k$ in which the $j^{th}$ entry equals $j$.  Then
$\pi_i^k = M_i^{\otimes k} \cdot {\bf{v}}^k$.

One useful observation pertains to the symmetry of our construction of
three permutations described in Section \ref{sec:construction}.
If we consider the set of permutations $\pi_1^k, \pi_2^k$ and
$\pi_3^k$, then the three permutations induced by $\{\pi_i^k\}$ on the set
of integers $[1,3^{k-1}]$ are isomorphic to the permutations
$\{\pi^{k-1}_i\}$.  This also holds for the permutations induced by
$\{\pi_i^k\}$ on $[3^{k-1}+1,2\cdot3^{k-1}]$ and to the permutations
induced by $\{\pi_i^k\}$ on $[2\cdot
  3^{k-1}+1, 3^k]$.

\begin{lemma}\label{iso}
Given permutations $\{\pi_i^k\}$, the three permutations induced on
$[1,3^{k-1}]$ (and on $[3^{k-1} +1, 2\cdot 3^{k-1}]$, $[2\cdot
  3^{k-1}+1, 3^k]$, respectively) are isomorphic to the permutations
$\{\pi^{k-1}_i\}$. 
\end{lemma}

\Proof The permutation $\pi_i^k$ is defined as $M_i^{\otimes k} \cdot
       {\bf v}^k$.  Note that this means that three copies of the
       permutation matrix corresponding to $\pi_i^{k-1}$ are placed in
       the three positions of matrix $M_i$, and all zero matrices of
       the same dimension are placed in the positions of $M_i$ that
       have a zero entry.  Thus, the same permutation, namely
       $\pi_i^{k-1}$, acts on each of the three following sets of
       integers: $[1,3^{k-1}]$, $[3^{k-1}+1, 2\cdot 3^{k-1}]$ and
       $[2\cdot 3^{k-1} +1, 3^k]$.  \qed

\section{Main Theorem}

Let ${\cal S}_k$ refer to the set system consisting of all prefixes of
our three permutations on $n = 3^k$ elements described in Section
\ref{sec:construction}.  Note that the set of all prefixes of the
permutations is a subset of all intervals of the permutations.  Since
we are proving a lower bound, it suffices to consider the set system
consisting only of prefixes.  Our main theorem is:
\begin{theorem}\label{main_thm}
$\mathrm{disc}({\cal S}_k) \geq \lceil{\frac{k}{3} + 1}\rceil = 
\lceil{\frac{\log_3{n}}{3} + 1}\rceil$.  
\end{theorem}

\section{Proof of Main Theorem}\label{ideas}

In our construction, as $k$ increases by 1, it is not necessarily the
case that the discrepancy increases by 1.  If this were true, then we
could prove a lower bound of $\log_3{n}$ rather than $\log_3{n}/3$.
However, one of our key ideas---roughly speaking---is that the {\em sum} of
the discrepancies of the set systems, each corresponding to one of the
permutations, increases by 1 as $k$ increases by 1.  We will use the
following definitions, which denote the maximum/minimum sum of the
prefixes of the set systems corresponding to each permutation for a
fixed coloring $\chi$:

\begin{eqnarray}
\text{disc}^k_{\text{L}^+}(\chi) & := & 
\max_{x, y, z \in [0,3^k]} \left(\chi(\alpha_1^k(x)) +
\chi(\alpha_2^k(y)) + \chi(\alpha_3^k(z)) \right),\\
\text{disc}^k_{\text{L}^-}(\chi) & := & 
\min_{x, y, z \in [0,3^k]} \left(\chi(\alpha_1^k(x)) +
\chi(\alpha_2^k(y)) + \chi(\alpha_3^k(z)) \right).
\end{eqnarray}
Although our set systems do not contain suffixes, we will also use the
following definitions:
\begin{eqnarray}
\text{disc}^k_{\text{R}^+}(\chi) & := & 
\max_{x, y, z \in [1,3^k+1]} \left(\chi(\omega_1^k(x)) +
\chi(\omega_2^k(y)) + \chi(\omega_3^k(z)) \right),\\
\text{disc}^k_{\text{R}^-}(\chi) & := & 
\min_{x, y, z \in [1,3^k+1]} \left(\chi(\omega_1^k(x)) +
\chi(\omega_2^k(y)) + \chi(\omega_3^k(z)) \right).
\end{eqnarray}
For a coloring $\chi:[3^k] \rightarrow \{-1, +1\}$, our goal is to
show the following:
\begin{eqnarray}
\text{disc}^{k}_{\text{L}^+}(\chi) & \geq & k + 3. \label{goal1}
\end{eqnarray}
Alternatively, if $\chi([3^k]) \leq -1$, then we want to show:
\begin{eqnarray}
\text{disc}^{k}_{\text{L}^-}(\chi) & \leq & -k - 3.\label{goal2}
\end{eqnarray}
This would imply our main theorem,
as one of the three set systems must
then have discrepancy at least $\lceil (k+3)/3 \rceil$.
However, we do not see how to directly use
\eqref{goal1} and \eqref{goal2} as an inductive hypothesis.  Thus, we
need a stronger inductive hypothesis, which is stated
in the following lemma and corollary.

\begin{lemma}\label{pos_left}
Let $\Delta = |\chi([3^k])|$.
If $\chi([3^k]) \geq 1$,
then: $${\mathrm{disc}}^k_{\mathrm{L}^+}(\chi),~
{\mathrm{disc}}^k_{\mathrm{R}^+}(\chi) ~\geq ~
k + \Delta + 2.$$
If $\chi([3^k]) \leq -1$,
then:
$$\mathrm{disc}^k_{\mathrm{L}^-}(\chi),~ \mathrm{disc}^k_{\mathrm{R}^-}(\chi) 
~ \leq ~ - k - \Delta -2.$$
\end{lemma}

Note that Lemma \ref{pos_left} implies our stated goal in
\eqref{goal1} and \eqref{goal2} and, therefore, our Main Theorem.
Indeed, since $3^k$ is odd, it must be the case for any coloring
$\chi:[3^k] \rightarrow \{-1, +1\}$ that $\Delta \geq 1$ and the theorem follows.  Before we prove
Lemma \ref{pos_left}, we show that Lemma \ref{pos_left} implies the
following corollary.

\begin{corollary}\label{pos_minus}
Let $\Delta = |\chi([3^k])|$.
If $\chi([3^k]) \leq -1$,
then: 
$$\mathrm{disc}^k_{\mathrm{L}^+}(\chi),~ \mathrm{disc}^k_{\mathrm{R}^+}(\chi) 
\geq k - 2\Delta + 2.$$
If $\chi([3^k]) \geq 1$,
$${\mathrm{disc}}^k_{\mathrm{L}^-}(\chi),~
{\mathrm{disc}}^k_{\mathrm{R}^-}(\chi) \leq
- k + 2\Delta -2.$$
\end{corollary}

\Proof Let us first consider the case in which $\chi([3^k]) \leq -1$.
Note that for each $\pi_i^k$, it is the case that for each $x \in
[0,3^k]$, $\chi(\alpha_i^k(x)) + \chi(\omega_i^k(x+1)) = \chi([3^k])$.
Therefore, for some coloring $\chi$, consider an $x \in [0,3^k]$ that
maximizes $\chi(\alpha_i^k(x))$.  Then $y = x+1$ is a value of $y
\in [1,3^k+1]$ that minimizes $\chi(\omega_i^k(y))$.
Thus, we have:
\begin{eqnarray}
\text{disc}^k_{\text{R}^-}(\chi) + \text{disc}^k_{\text{L}^+}(\chi) &
= & 3\chi([3^k]) \quad \Rightarrow\\
\text{disc}^k_{\text{L}^+}(\chi) & = & 3\chi([3^k]) - 
\text{disc}^k_{\text{R}^-}(\chi).
\end{eqnarray}
By Lemma \ref{pos_left}, we have:
\begin{eqnarray}
\text{disc}^k_{\text{L}^+}(\chi) 
& \geq & -3\Delta  + k + \Delta + 2 \\
& = & k - 2\Delta + 2.
\end{eqnarray} 
An analogous argument works to give the same lower bound on
$\text{disc}^k_{\text{R}^+}$ when $\chi([3^k]) \leq -1$.
Now consider the case in which $\chi([3^k]) \geq 1$.  We have:
\begin{eqnarray}
\text{disc}^k_{\text{R}^+}(\chi) + \text{disc}^k_{\text{L}^-}(\chi) &
= & 3\chi([3^k]) \quad \Rightarrow\\
\text{disc}^k_{\text{L}^-}(\chi) & = & 3\chi([3^k]) - 
\text{disc}^k_{\text{R}^+}(\chi).
\end{eqnarray}
By Lemma \ref{pos_left}, we have:
\begin{eqnarray}
\text{disc}^k_{\text{L}^-}(\chi) 
& \leq & 3\Delta  - k - \Delta - 2\\
& = & - k + 2\Delta - 2.
\end{eqnarray} 
The argument for the upper bound on $\text{disc}^k_{\text{R}^-}$ when
$\chi([3^k]) \geq 1$ is symmetric.
\qed

\subsection{Proof of Lemma 2}

Now we will prove Lemma \ref{pos_left} using induction.  

\subsubsection*{Base Case: $k = 1$}

Suppose that $\chi([3]) \geq 1$.  There are only two possibilities for
such colorings:
\begin{eqnarray}
\begin{pmatrix}
1 & -1 & 1 \\
1 & 1 & -1 \\
-1 & 1 & 1
\end{pmatrix} \quad \quad
\begin{pmatrix}
1 & 1 & 1 \\
1 & 1 & 1 \\
1 & 1 & 1
\end{pmatrix}
\end{eqnarray}
Suppose $\chi([3]) = 1$.  The only way to achieve such a coloring is
to have two of the elements be colored `$+1$' and one element be
colored `$-1$'.  Then one of the permutations has a prefix (suffix)
with value two, while each of the other two permutations have prefixes
(suffixes) with value one.  Thus, we have:
$\text{disc}^1_{\text{L}^+}(\chi), \text{disc}^1_{\text{R}^+}(\chi) =
4 \geq k + \Delta + 2 = 4$.  Now
suppose that $\chi([3]) = 3$.  In this case, each permutation has a
prefix (suffix) with value three.  Thus,
$\text{disc}^1_{\text{L}^+}(\chi), \text{disc}^1_{\text{R}^+}(\chi) =
9 \geq k + \Delta + 2 = 6$.  Thus, Lemma \ref{pos_left} holds for $\chi([3]) \geq 1$
when $k=1$.

When $\chi([3]) = -1$, the same arguments can be used to show that
$\text{disc}^1_{\text{L}^-}(\chi), \text{disc}^1_{\text{R}^-}(\chi) =
-4 \leq -k -\Delta -2 = -4$.  Similarly, when $\chi([3]) = -3$,
$\text{disc}^1_{\text{L}^-}(\chi), \text{disc}^1_{\text{R}^-}(\chi) =
-9 \leq -6$.  This concludes the proof of the base case.

\subsubsection*{Inductive Step}

Now we assume that the Lemma and its Corollary are true for
$k-1$ and prove the Lemma (and thus, the Corollary) true for $k$.

For some fixed $\chi: [3^k] \rightarrow \{-1, +1\}$, let $a$, $b$ and
$c$ denote the values of the three blocks of $3^{k-1}$ consecutive
integers in the recursive construction, i.e. $\chi([1,3^{k-1}]),
\chi([3^{k-1}+1, 2\cdot 3^{k-1}])$ and $\chi([2\cdot 3^{k-1}+1,
  3^k])$, although not necessarily in this order.  We always assume
that $a \geq b \geq c$, i.e. the value of the block with the largest
value is denoted by $a$, etc.  Note that $a, b$ and $c$ are each odd
numbers, because they always represent the values of intervals with
odd length.  Each permutation in $\{\pi^k_i\}$ corresponds to some
permutation of $a, b$ and $c$.  Without changing the discrepancy, we
can rearrange the three permutations to form one of the following two
configurations, in which each row corresponds to one of the three
permutations in $\{\pi_i^k\}$.
\begin{eqnarray*}
\text{(I)}~
\begin{pmatrix}
a & b & c \\
c & a & b \\
b & c & a
\end{pmatrix} =
\begin{pmatrix}
b & c & a\\
a & b & c \\
c & a & b 
\end{pmatrix}, \quad \quad
\text{(II)}~
\begin{pmatrix}
a & c & b \\
b & a & c \\
c & b & a
\end{pmatrix}.
\end{eqnarray*}

First we consider the case in which $\chi([3^k]) \geq 1$.  This
implies that $a + b + c \geq 1$.  
There are two subcases:
\begin{itemize}
\item[(i)] $a \geq b \geq 1$ (and $c \geq 1$ or $c \leq -1$), 

\item[(ii)]$a \geq 1$ and $c \leq b \leq -1$.
\end{itemize}

First, we consider case (i) and configuration (I).  If we look at a
permutation of the rows so that the blocks with value $b$ are on the
diagonal (as shown), then in configuration (I), the value of the
blocks below the diagonal are positive (which is desirable).  Thus, we
can consider the three prefixes corresponding to the permutations of
the block with value $b$.  Suppose, without loss of generality (and
for ease of notation) that the block with value $b$ is $[1, 3^{k-1}]$.
In this case, the permutations on the diagonal are $\pi_1^{k-1},
\pi_2^{k-1}$ and $\pi_3^{k-1}$.  By the inductive assumption, for any
$\chi:[3^{k-1}] \rightarrow \{-1, +1\}$, there are three corresponding
prefixes $\alpha^{k-1}_1(x), \alpha^{k-1}_2(y)$ and
$\alpha^{k-1}_3(z)$, for some integers $x_1,x_2,x_3 \in [0,3^{k-1}]$, such
that:
\begin{eqnarray}
\chi(\alpha^{k-1}_1(x_1)) +
\chi(\alpha^{k-1}_2(x_2)) + 
\chi(\alpha^{k-1}_3(x_3)) & = &  
\text{disc}^{k-1}_{\text{L}^+}(\chi) \\
& \geq & (k-1) + b + 2. \label{ind_assump}
\end{eqnarray}
Note that if either the block $[3^{k-1}+1, 2\cdot 3^{k-1}]$ or the
block $[2\cdot 3^{k-1}+1, 3^k]$ had value $b$, and therefore appeared
on the diagonal of configuration (I), then by Lemma \ref{iso}, we
see that these permutations are isomorphic to $\{\pi_i^{k-1}\}$.  This
allows us to use the inductive hypothesis in these cases as well, and
to draw the same conclusion as we drew in \eqref{ind_assump}.

Now we consider some $\chi:[3^k] \rightarrow \{-1, +1\}$.  This
coloring induces a coloring on $[3^{k-1}]$ for which the above
assumption in \eqref{ind_assump} holds.  Suppose that $\pi^{k-1}_h,
\pi^{k-1}_{j}$ and $\pi^{k-1}_{\ell}$, for $h,j, \ell \in \{1,2,3\}$,
correspond to the permutations of block $[3^{k-1}]$ that appear in the
first, second and third rows of the configuration, respectively.  For
the fixed coloring $\chi$ on $[3^k]$, our goal is to show that there
are three prefixes of the three permutations $\{\pi_i^k\}$ such that
we can lower bound the value of the sum of these prefixes with respect
to the fixed coloring $\chi$.  The prefix of the permutation
corresponding to the first row of the configuration is
$\alpha_h^{k-1}(x_h)$.  For the permutation corresponding to the second
row of the configuration, we add the block with value $a$ to the front
of $\alpha_j^{k-1}(x_j)$.  For the permutation corresponding to the
third row of the configuration, we add the block with value $a$ to the
front of $\alpha_{\ell}^{k-1}(x_{\ell})$ preceded by the block with value
$c$.
Thus, by the inductive
hypothesis, we have that:
\begin{eqnarray}
\text{disc}^k_{\text{L}^+}(\chi) & \geq &
\chi(\alpha_h^{k-1}(x_h)) + \left(a + \chi(\alpha_j^{k-1}(x_j))\right) + \left(c + a + \chi(\alpha_{\ell}^{k-1}(x_{\ell}))\right)\\
& = &
\text{disc}^{k-1}_{\text{L}^+}(\chi) + 2a + c\\
& \geq & (k - 1) + b + 2 + 2a + c \label{oneone}\\
& \geq & k + \Delta + 1 + a\\
& \geq & k + \Delta + 2.
\end{eqnarray}
The last inequality follows from the fact that in case (i),
$a \geq 1$.  Thus, the inductive step holds for case (i),
configuration (I).

Now let us consider configuration (II).  In this
case, we consider a permutation of the rows so that the blocks with
value $a$ occupy the diagonal.  By the same reasoning as discussed
previously and by induction, we have:
\begin{eqnarray}
\text{disc}^k_{\text{L}^+}(\chi) & \geq &
\text{disc}^{k-1}_{\text{L}^+}(\chi) + 2b + c \\
& \geq & (k-1) + a + 2 + 2b + c\\
& \geq & k + \Delta + b + 1\\
& \geq & k + \Delta + 2.
\end{eqnarray}
Since in case (i), $b \geq 1$, the inductive step holds for case (i),
configuration (II).

Now we consider case $(ii)$, when $a \geq 1$ and $c \leq b \leq -1$.
In this case, we again have the above two configurations:
\begin{eqnarray}
\text{(I)}~
\begin{pmatrix}
a & b & c \\
c & a & b \\
b & c & a
\end{pmatrix} = 
\begin{pmatrix}
b & c & a\\
a & b & c \\
c & a & b 
\end{pmatrix}, \quad \quad
\text{(II)}~
\begin{pmatrix}
a & c & b \\
b & a & c \\
c & b & a
\end{pmatrix} =
\begin{pmatrix}
c & b & a\\
a & c & b \\
b & a & c 
\end{pmatrix}.
\end{eqnarray}
Note that in case (ii), for both configurations (I) and (II), we 
use Corollary \ref{pos_minus}.  We consider configuration (I) first.
\begin{eqnarray}
\text{disc}^k_{\text{L}^+}(\chi) & \geq &  \text{disc}^{k-1}_{\text{L}^+}(\chi) + 2a +
c \\
& \geq & (k - 1) -2|b| + 2 + 2a + c\\
& \geq & (k - 1) + 2b + 2 + 2a + c\\
& \geq & k + \Delta + a + b + 1\\
& \geq & k + \Delta + 2. 
\end{eqnarray}
Since we have $a + b + c \geq 1$, it follows that $a + b \geq 1 - c
\geq 2$.  Thus, case (ii) holds for configuration (I).  Now let us
consider configuration (II).  We have:
\begin{eqnarray}
\text{disc}^k_{\text{L}^+}(\chi) & \geq &  \text{disc}^{k-1}_{\text{L}^+} + 2b +
c \\
& \geq & (k - 1) -2|c| + 2 + 2a + b\\
& \geq & (k - 1) + 2c + 2 + 2a + b \\
& \geq & k + \Delta + a + c  + 1\\
& \geq & k + \Delta + 2.
\end{eqnarray}
Since we have $a + b + c \geq 1$, it follows that $a + c \geq 1-b \geq
2$.  Thus, case (ii) holds for configuration (II).

The proof of the lower bound on $\text{disc}^k_{\text{R}^+}(\chi)$ is
symmetric to the one we have just given for
$\text{disc}^k_{\text{L}^+}(\chi)$.  Instead of adding the blocks
whose values lie in the lower left hand triangle to form the new
prefixes, we use the blocks whose values lie in the upper right hand
triangle.

Finally, we need to show that
if $\chi([3^k]) \leq -1$,
then:
\begin{eqnarray}
\mathrm{disc}^k_{\mathrm{L}^-}(\chi),~ \mathrm{disc}^k_{\mathrm{R}^-}(\chi) 
~ \leq ~ - k - \Delta -2. \label{to_prove}
\end{eqnarray}
Note that this follows from our proof of the first part of Lemma
\ref{pos_left}, namely that 
when $\chi([3^k]) \geq 1$,
then:
\begin{eqnarray}
\mathrm{disc}^k_{\mathrm{L}^+}(\chi),~ \mathrm{disc}^k_{\mathrm{R}^+}(\chi) 
~ \geq ~ k + \Delta +2.\label{proved}
\end{eqnarray}
This is due to the observation that if consider a coloring $\chi:[3^k]
\rightarrow \{-1, +1\}$ such that $\chi([3^k]) \leq -1$, and it is the
case that \eqref{to_prove} does {\em not} hold, then consider
$\chi^- = - \chi$, i.e. the negation of $\chi$.  It follows that
$\chi^-([3^k]) \geq 1$, but \eqref{proved} does not hold for coloring
$\chi^-$, which is a contradiction.\qed

\section{Discussion}

Our construction gives only a single set of three permutations for
each value of $k$.  However, our construction can actually generate up
to $2^k$ sets of permutations for each $k$.  The construction we have
described in this paper can be viewed as taking one right shift for
each set of blocks in the second permutation and two right shifts for
each set of blocks in the third permutation.  However, for each $h: 1
\leq h \leq k$, we can choose right or left, thus generating many more
sets of permutations.  Because of the symmetry of our proofs, they
should still hold for these constructions as well.  This observation
was made by Ofer Neiman.

\section{Acknowledgements}

We thank Moses Charikar and Ofer Neiman for
helpful discussions.

{\small
\bibliography{beck-new}}

\begin{thebibliography}{EPR11}

\bibitem[Ban10]{DBLP:conf/focs/Bansal10}
Nikhil Bansal.
\newblock Constructive algorithms for discrepancy minimization.
\newblock In {\em FOCS}, pages 3--10. IEEE Computer Society, 2010.

\bibitem[Boh90]{DBLP:journals/rsa/Bohus90}
G{\'e}za Bohus.
\newblock On the discrepancy of 3 permutations.
\newblock {\em Random Struct. Algorithms}, 1(2):215--220, 1990.

\bibitem[EPR11]{DBLP:conf/soda/EDR11}
Friedrich Eisenbrand, D{\"o}m{\"o}t{\"o}r P{\'a}lv{\"o}lgyi, and Thomas
  Rothvo{\ss}.
\newblock Bin packing via discrepancy of permutations.
\newblock In {\em SODA}, pages 1029--1034, 2011.

\bibitem[KK82]{DBLP:conf/focs/KarmarkarK82}
Narendra Karmarkar and Richard~M. Karp.
\newblock An efficient approximation scheme for the one-dimensional bin-packing
  problem.
\newblock In {\em FOCS}, pages 312--320. IEEE, 1982.

\bibitem[Mat10]{matousek2010geometric}
J.~Matousek.
\newblock {\em Geometric Discrepancy: An Illustrated Guide}.
\newblock Springer Verlag, 2010.

\bibitem[Spe87]{spencer-lectures}
Joel Spencer.
\newblock {\em Ten Lectures on the Probabilistic Method}.
\newblock Society for Industrial and Applied Mathematics, 1987.

\bibitem[SST01]{SST}
J.~H. Spencer, A.~Srinivasan, and P.~Tetali.
\newblock The discrepancy of permutation families, 2001.

\end{thebibliography}

\end{document}